\documentclass[twocolumn,aps,prl,showpacs,showkeys,
tightenlines]{revtex4-1}
\usepackage{xcolor}
\usepackage{graphicx}
\usepackage{float}
\usepackage{multirow}
\usepackage{epsfig}
\usepackage{tabularx}

\begin{document}

\title{Neutron radii
and neutron skin of neutron-rich nuclei
deduced from proton-nucleus total reaction cross sections }

\author{I. A. M. Abdul-Magead, Eman Hamza, Badawy Abu-Ibrahim}
\affiliation{Department of Physics, Cairo University,
Giza 12613, Egypt}


\begin{abstract}
 A new method is proposed to deduce the neutron radii of
neutron-rich nuclei. This method requires measuring the reaction cross sections
of both the neutron-rich nucleus and its stable
isotope at the same energy on a proton target. Using this  method and the available experimental data
of $^{12,22}$C + $p$ and $^{9,14}$Be + $p$ at 40$A$ MeV,
the neutron radii of $^{22}$C and $^{14}$Be  have been deduced for the first time.

\end{abstract}

\maketitle

Neutron halo structure has been observed among the
neutron-rich nuclei, it is manifested by the neutron
density extending to extraordinary large radial distance.
The neutron halo structure is  indicated first by an increase
in the root-mean-square point matter radii (matter radii) \cite{tanihata85}.
The matter radii ($ {\bar r}_{m}=\langle r^{2}_{m} \rangle^{1/2}$)
of unstable nuclei are usually calculated by
measuring their reaction (interaction)
cross sections on Carbon target \cite{tanihata85}
or proton target \cite{tanaka10} by the  Glauber theory.

The root-mean-square point neutron radii (neutron radii)
play an  important role
in nuclear physics and astrophysics.
Neutron radii ($ {\bar r}_{n}=\langle r^{2}_{n} \rangle^{1/2}$)
are essential to extract the neutron skin thickness
($S_{n}= {\bar r}_{n}- {\bar r}_{p}$ ).
The neutron skin thickness is strongly correlated to
 the radius of the low-mass neutron stars and to the density dependence
of the symmetry energy~\cite{fattoyev18,sagawa18},
(often encoded in a quantity denoted by $L$)  which is related to the
pressure of the pure neutron matter at the saturation density~\cite{iida03}.
The origin of the size of both the thickness of the neutron skin
and the neutron star is the pressure of the neutron-rich matter,
 pushing either against surface tension in an atomic nucleus
or against gravity in a neutron star.
The stellar radius and the neutron skin are sensitive to the
same equation of state~\cite{fattoyev18}.
Recently, neutron star physics has received renewed  interest  since
the LIGO-Virgo collaboration made the first direct detection of gravitational waves
from the coalescence of a neutron star binary system~\cite{abbott17}. The analysis of the data placed constraints on the tidal effects of the coalescing bodies, which were then translated to constraints on neutron star radii~\cite{abbott18}.
Also, the knowledge of the nuclear symmetry energy is relevant for
the Standard Model tests via atomic parity violation~\cite{sil}.

Electron-nucleus scattering and measurement of muonic x rays are
used to determine the charge radii of stable nuclei. These techniques
cannot be used for neutron-rich isotopes due to the weak beam intensities.
High precision measurements of the charge radii
($ {\bar r}_{c}=\langle r^{2}_{c} \rangle^{1/2}$)
of He~\cite{mueller07,brodeur12},
Li~\cite{sanchez06},
Be~\cite{nortershauser09,krieger12}, and Mg~\cite{Yordanov12} isotopes
have been achieved using the isotope-shift
technique.
Combining the isotope-shift data with
the data deduced from the reaction (interaction) cross section,
one can obtain the neutron radii
of the neutron-rich isotopes from the relation
${\bar r}^{2}_{n} = (A/N) {\bar r}^{2}_{m} - (Z/N) {\bar r}^{2}_{p}$
and hence their neutron skin thickness~\cite{Suzuki95}.

Due to the limitation of the
isotope-shift technique, mainly due to the low luminosity of rare isotopes
close to the neutron drip line,
the measurement of the charge-changing cross section ($\sigma_{cc}$)
has been used to determine the proton radii of
neutron-rich Carbon-isotopes~\cite{yamaguchi11}.
The authors of Ref.~\cite{yamaguchi11} report that they scaled the experimental cross section
to reproduce the known charge radius of $^{12}$C.
They also combined the charge-changing
cross section data with the reaction cross
section measurements to obtain the neutron radii of $^{15,16}$C,
and hence their neutron skins.
Refs.~\cite{estrade,kanungo} used charge-changing cross section to deduce
the proton radii of $^{12-17}$B and $^{12-19}$C without any scaling factor.

The knowledge of the neutron radii of stable nuclei is scarce compared to that
of proton radii,  its main sources are  parity-violation in
electron scattering~\cite{abrahamyan}
and hadron scattering experiments, see for example~\cite{clark}.
Currently, these methods cannot be used for neutron-rich nuclei.
In this letter,  a new method is proposed
in order to deduce the neutron radii
of neutron-rich nuclei.
The method requires only measuring the difference between
the reaction cross sections
of the studied neutron-rich nucleus and its
stable isotope on proton targets at the same energy.    Also, the efficiency of this
method is demonstrated and it is applied to determine the neutron radii of $^{22}$C and $^{14}$Be.


The total reaction cross section of a nucleus ($A=p+n$) incident on
a proton is expressed in the Optical Limit Approximation (OLA)
of Glauber theory \cite{glauber} as
{\bm b}
\begin{equation}\label{gfn}
  \sigma^{ \rm R}_{\rm A} = \int{d {\bm \bf b}\,
  \left(1-{\rm e}^{-2{\rm Im}[\chi_{p}({\bm \bf b})+\chi_{n}({\bm\bf b})]}\right)},
\label{reaccs}
\end{equation}
where ${\bm\bf b}$ is the impact parameter vector perpendicular
to the beam (z) direction, and
$\chi({\bm\bf b})$ is the phase-shift function defined as
$i\chi_{N}( {\bm\bf b} ) = -{\int{d{\bm \bf s} }} \rho_{N}({\bm \bf s})
\Gamma_{pN}({\bm \bf s} + {\bm \bf b})$,
where $\chi_{N}$ implies the phase shift due to the
protons ($N=p$) or neutrons ($N=n$) inside the nucleus.
The function $\rho_{N}({\bm s})$ is the z-integrated density distribution
of the protons (neutrons).
The finite-range profile function, $\Gamma_{pN}$,
for $pp$ and $pn$ scatterings, is usually parameterized in the form
$\Gamma_{pN}({\bm \bf b}) =
\frac{1-i\alpha_{pN}}{4\pi\beta_{pN}}\,\,
\sigma_{pN}^{\rm tot}\, {\rm e}^{-{\bm \bf b}^2 /(2\beta_{pN}) }$,
where $\alpha_{pN}$ is the ratio of
the real to the imaginary part of the
$pp$ ($pn$) scattering amplitude in the forward direction,
$\sigma_{pN}^{\rm tot}$ is the $pp$ ($pn$) total cross section,
and $\beta_{pN}$ is the slope parameter of the
$pp$ ($pn$) elastic scattering differential cross section.
For the zero-range approximation, it is parameterized in the form
$\Gamma_{pN}({\bm \bf b}) =
\frac{1-i\alpha_{pN}}{2}\,\,
\sigma_{pN}^{\rm tot}\, \delta({\bm \bf b})$.
The validity of the OLA has already been tested with
stable and unstable nuclei incident on a proton target~\cite{badawy08,nagahisa}.

The reaction cross section shift
between a neutron-rich nucleus and its stable isotope is defined as
$\delta\sigma^{\rm R}_{\rm A_{2},\rm A_{1}}=\sigma^{\rm R}_{\rm A_{2}}-\sigma^{\rm R}_{A_{1}}$.
Both $\sigma^{\rm R}_{\rm A_{2}}$ and $\sigma^{\rm R}_{\rm A_{1}}$
are measured on a proton target and at the same energy.
From Eq.~(\ref{reaccs}), the reaction cross section shift is given by
\begin{eqnarray}
 &&  \delta\sigma^{\rm R}_{\rm A_{2},\rm A_{1}}=
 2\pi\int{b}\,
  {\rm e}^{-2{\rm Im}[\chi_{p_{1}}({\bm b})+\chi_{n_{1}}({\bm b})]}\nonumber\\
 && \times\left(1-{\rm e}^{-2{\rm Im}[\Delta\chi_{p}({\bm b})+
 \Delta\chi_{n}({\bm b})]}\right)db,
\label{dreac}
\end{eqnarray}
where, $\Delta\chi_{N}({\bm b})=
\chi_{N_{2}}({\bm b})-\chi_{N_{1}}({\bm b})$
and
\begin{eqnarray}
&&-2{\rm Im}[\Delta\chi_{N}({\bm b})] = -\,
\frac{\sigma_{pN}^{\rm tot} }{2\pi\beta_{pN}}\nonumber\\
&&\times\int{d{\bm s}}\,[\rho_{N_{2}}({\bm s})- \rho_{N_{1}}({\bm s})]
{\rm e}^{-({\bm b}+{\bm s})^2 /(2\beta_{pN})}.
\label{chmpf}
\end{eqnarray}
For the zero range approximation, it reduces to
\begin{equation}
-2{\rm Im}[\Delta\chi_{N}({\bm b})] =
 -\,\,
\sigma_{pN}^{\rm tot}\, (\rho_{N2}({\bm b})-\rho_{N1}({\bm b})).
\label{chmn}
\end{equation}
It has been shown, in Ref.~\cite{kaki}, that various nuclear density distributions
having the same matter radii provide almost the same values of the
$p$-nucleus reaction cross section in the energy region 100$A$-800$A$ MeV.
Furthermore, charge radii obtained from the isotope-shift technique
show that the maximum
difference between the measured charge radii of the isotopes of the same element
for $^{4,6,8}$He, $^{6,8,9,11}$Li, $^{9-11}$Be, and $^{24-32}$Mg
are 0.392$\pm$0.019 fm (23$\%$ from that of $^{4}$He)~\cite{mueller07},
0.3$\pm$0.065 fm (12$\%$ from that of $^{6}$Li)~\cite{sanchez06},
0.162$\pm$0.03 fm (6$\%$ from that of $^{9}$Be)~\cite{nortershauser09}, and
0.158$\pm$0.008 fm (5$\%$ from that of $^{24}$Mg)~\cite{Yordanov12}, respectively.
Also, the proton radii obtained from the charge-changing cross section indicate that
the maximum difference between the measured proton radii of the isotopes
of the same element for
$^{12-17}$B, $^{12-16}$C, and $^{12-19}$C, are
0.35$\pm$0.04 fm (15$\%$ from that of $^{10}$B)~\cite{estrade},
0.09$\pm$0.012 fm,(4$\%$ from that of $^{12}$C)~\cite{tran}, and
0.12$\pm$0.08 fm (5$\%$ from that of $^{12}$C)~\cite{kanungo}, respectively.
As one sees, the maximum difference
between the proton radii for the
isotopes of the same element is less than 10$\%$ except for He, Li, and B isotopes.
  It has been shown in Ref.~\cite{kanungo} that the effect of the center-of-mass
motion of the halo on the proton radius becomes smaller with increasing mass number.
The values of the charge radii, given in~\cite{adndt},
show that the differences between the charge radii of the
isotopes of the same element are small.
Based on the above discussion,  we assume that the contribution of
$\Delta\chi_{p}({\bm b})$ to the value of the
reaction cross section shift is negligible.
This means that $\Delta\chi_{p}({\bm b}) \simeq 0$.
Under this assumption,  Eq.~(\ref{dreac}) reduces to
\begin{eqnarray}
 &&\delta\sigma^{\rm R}_{\rm A_{2},\rm A_{1}} \simeq
 2\pi\int{b}\,
  {\rm e}^{-2{\rm Im}[\chi_{p1}({\bm b})+\chi_{n1}({\bm b})]}\nonumber\\
 && \times\left(1-{\rm e}^{-2{\rm Im}[\Delta\chi_{n}({\bm b})]}\right)db.
\label{dreac1}
\end{eqnarray}
Note that Eq.~(\ref{dreac1}) does not depend on the proton
density of the neutron-rich isotope.
In  Eq.~(\ref{dreac1}), the neutron and proton
density distributions of the stable isotope can be obtained from
literature eg. \cite{devries,patterson},
while the neutron density distribution of the neutron-rich isotope is the key quantity.
To check the validity of the approximation in Eq.~(\ref{dreac1}),
 the values of $\delta\sigma^{\rm R}_{\rm A_{2},\rm A_{1}}$
  calculated using Eq.~(\ref{dreac1}) is  compared to the exact values obtained from
Eq.~(\ref{dreac}). Thus, the factor FS is defined as
\begin{equation}
{\rm FS}=\frac{\delta\sigma^{\rm R}_{\rm A_{2},\rm A_{1}}[{\rm Eq.}~(\ref{dreac})]}
{\delta\sigma^{\rm R}_{\rm A_{2},\rm A_{1}}[{\rm Eq.}~(\ref{dreac1})]},
\end{equation}
and its values are calculated for Be, B, C, O, Ne and Mg isotopes at different projectile energies.
In the present calculations, the harmonic oscillator (HO) density~\cite{badawyJPSJ} has been used for
Be and B with radii taken from~\cite{krieger12} and \cite{tel}(SKM$^{\star}$), respectively.
The densities of Carbon isotopes are taken from Ref.~\cite{badawy08}.
 For O, Ne and Mg isotopes, the Hartree-Fock densities have been used~\cite{ripl}.
The nucleon-nucleon interaction parameters are taken from Ref.~\cite{badawy08}.

\begin{figure}[t]
\hspace{-1cm}\epsfig{file=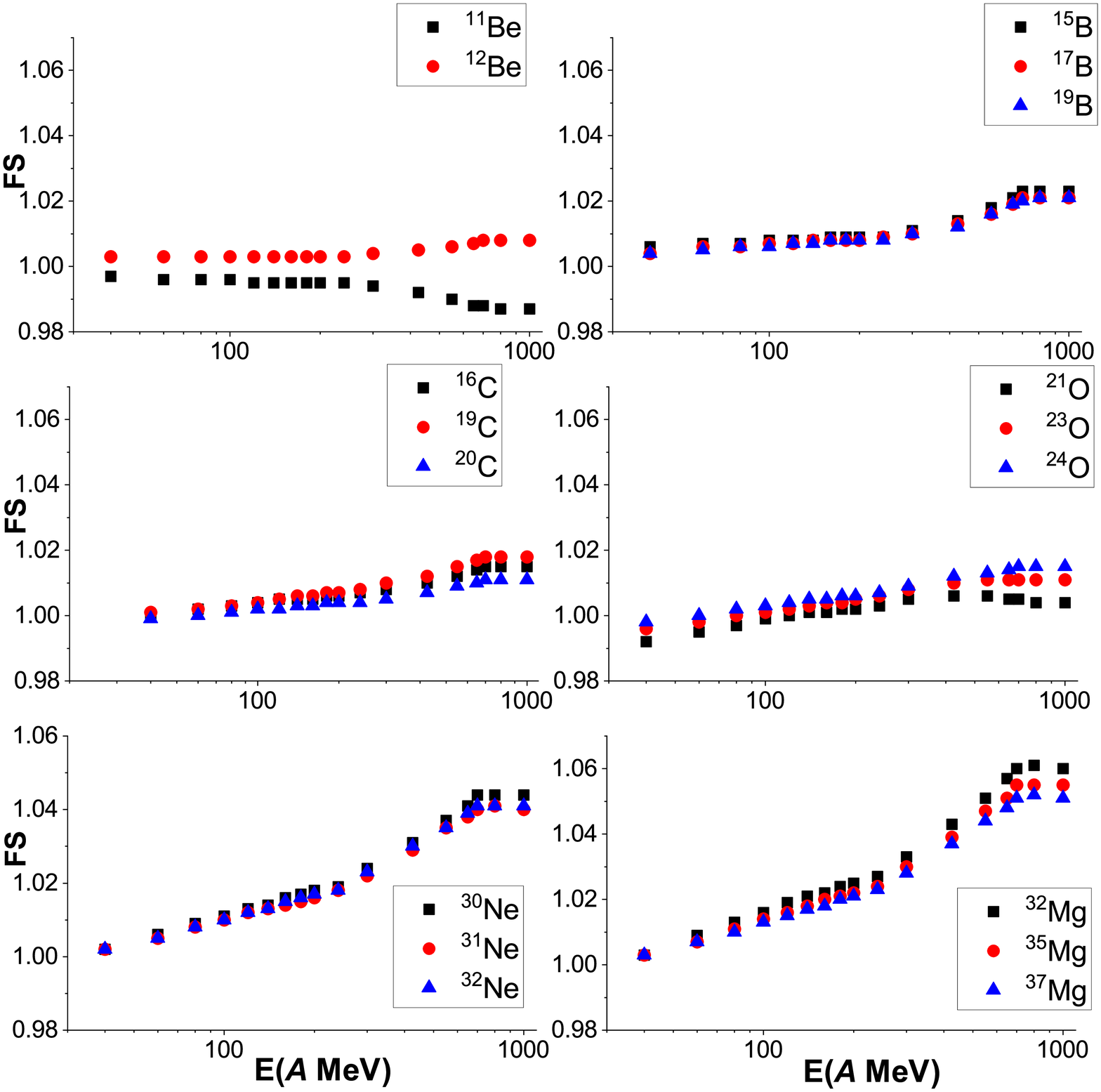,scale=0.32}
\caption{The FS factor as a function of the energy
for Be, B, C, O, Ne and Mg isotopes}
\label{fig1}
\end{figure}

A sample of  the results is given in Fig.~\ref{fig1},
 showing the FS factor as a function of energy for $^{11,12}$Be, $^{15,17,19}$B,
$^{16,19,20}$C, $^{21,23,24}$O, $^{30,31,32}$Ne and $^{32,35,37}$Mg
where the reaction cross section shifts are calculated
with respect to the stable nuclei
$^9$Be, $^{10}$B, $^{12}$C, $^{16}$O, $^{20}$Ne and $^{24}$Mg respectively.
As  can be seen from the figure, the deviation between the values
calculated using Eq.~(\ref{dreac1}) and the exact ones calculated using Eq.~(\ref{dreac})
does not exceed  $7\%$ for all the considered isotopes in the considered  energy range.

\begin{figure}[t]
\epsfig{file=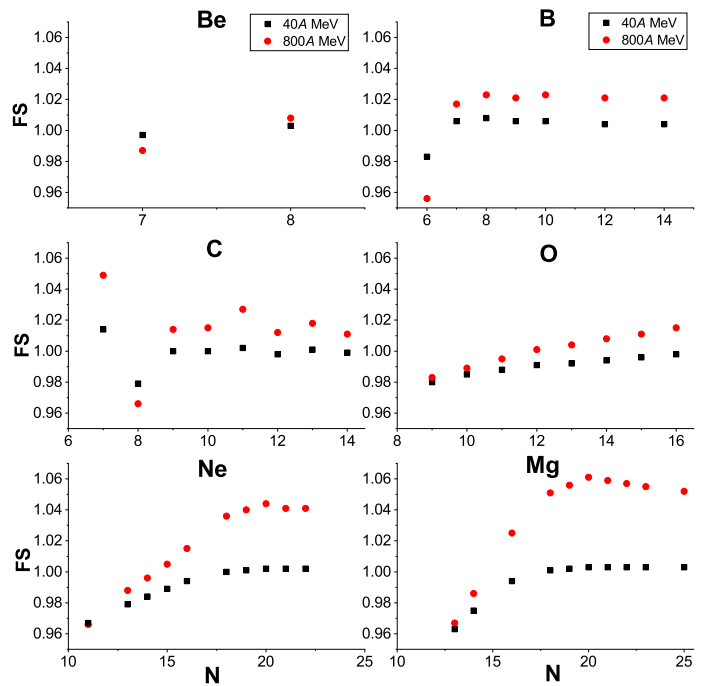,scale=0.32}
\caption{The factor FS as a function of the neutron number of Be, B, C, O, Ne and Mg
isotopes at 40$A$ MeV and 800$A$ MeV.}
\label{isotopes}
\end{figure}

Figure \ref{isotopes} shows the FS factor as a function of the neutron number
of all the considered isotopes at the energies 40$A$ MeV and 800$A$ MeV.
Confirming that the deviation between the values calculated
using Eq.~(\ref{dreac1}) and the exact ones calculated using Eq.~(\ref{dreac})
does not exceed $7\%$ irrespective of the neutron number.

\begin{figure}[t]
\epsfig{file=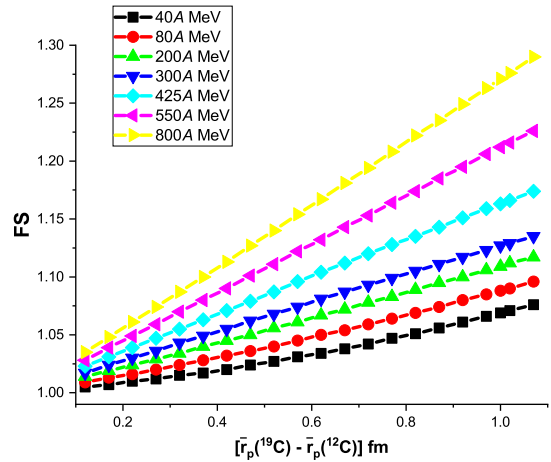,scale=0.3}
\caption{The factor FS as a function of the difference between
proton radii of $^{19}$C and $^{12}$C  at energies 40$A$-800$A$ MeV.}
\label{figrmsp}

\end{figure}

In order to examine the effect of the increase of the proton radii of the neutron-rich nuclei on the validity of the  approximation used in Eq.~(\ref{dreac1}),
a model analysis with HO density distribution for $^{12,19}$C has been preformed.
The configuration of the $^{12}$C wave function is assumed to be
$(0s_{1/2})^{2}(0p_{3/2})^{4}$ for both protons and neutrons.
The HO length parameter is fixed in such a way to reproduce the
proton radius, 2.33 fm, extracted from the charge radius.
Also,  the neutron radius of $^{19}$C  is fixed to be 3.37 fm~\cite{badawy08},
with the configuration
$(0s_{1/2})^{2}(0p_{3/2})^{4}(0p_{1/2})^{2}(0d_{5/2})^{4}(1s_{1/2})^{1}$.
 The HO length parameter is increased in order to
increase the proton radius of $^{19}$C from 2.33 fm by 0.05 fm in each step.
Figure \ref{figrmsp} shows the relation between
the FS factor and the difference in proton radii
of $^{19}$C and $^{12}$C at different energy values.
If the proton radius of $^{19}$C increases by almost 10$\%$, 20$\%$, 30$\%$
from that of $^{12}$C, the FS factor increases by almost
1$\%$ (6$\%$), 3$\%$ (13$\%$), 5$\%$ (20$\%$), respectively
at  incident energy 40 MeV (800 MeV).
From this figure, one notices that the approximation given in Eq.~(\ref{dreac1})
is very effective at energies less than 300$A$ MeV.

As a first application of the proposed method,
 the neutron radius of $^{22}$C is determined.
The $^{22}$C nucleus is known as a two-neutron halo type.
It has been studied in the three-body model of
$^{20}$C+n+n predicting a binding energy in the range 0.122-0.489 MeV
and a matter radius (neutron radius) 3.61-4.11 fm (3.96-4.58 fm)~\cite{badawy07}.
Accurate experimental data on $^{22}$C binding energy is not yet available,
so $^{22}$C is still attracting attention~\cite{shulgina18}.
The interaction cross section of $^{22}$C on a proton target at 40$A$ MeV was
measured first by Tanaka {\it et al.}~\cite{tanaka10},  in 2010,
to be $\sigma^{\rm R}_{22}$[Exp]=1338 $\pm$274 mb,
resulting in the rather large matter radius
5.4 $\pm$ 0.9 fm.
On the other hand, Togano {\it et al.}~\cite{togano16}, in 2016,  measured the interaction cross section
of $^{22}$C on a $^{12}$C target at almost 240$A$ MeV
obtaining a matter radius 3.44 $\pm$ 0.08 fm.
Nevertheless,  the results of Ref.~\cite{tanaka10} have been used  in the present calculations
since the  method requires a proton target.
The available data of the reaction cross section of p+$^{12}$C at 40 MeV is for natural Carbon and  has the value of 371$\pm$11 mb~\cite{menet}.
Thus, the reaction cross section shift for $^{22}$C and $^{12}$C incident
on a proton target at 40$A$ MeV is $\delta\sigma^{\rm R}_{22,12}$[Exp]=967$\pm$285 mb.

\begin{figure}[t]
\hspace{-0.5cm}\epsfig{file=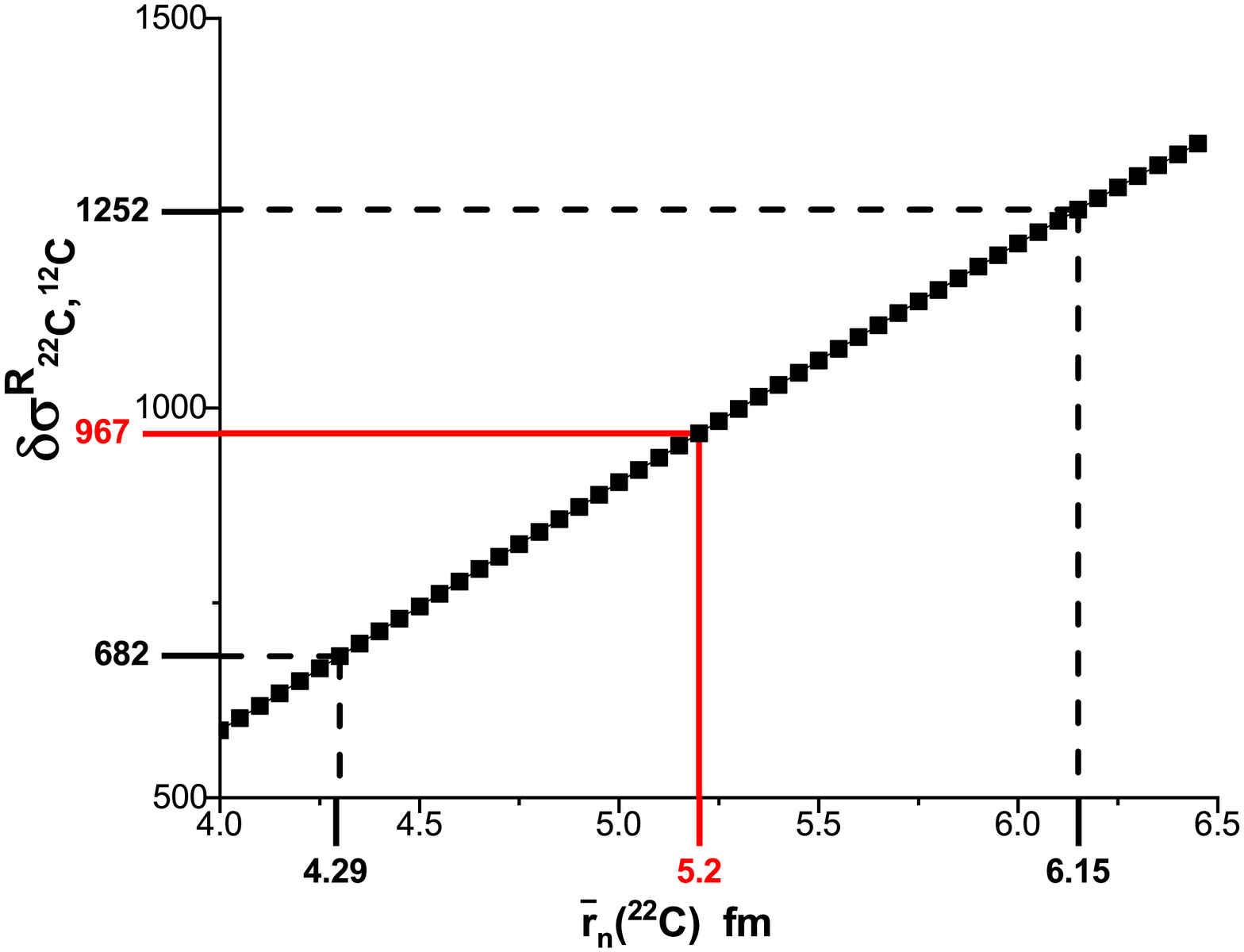,scale=0.3}
\caption{The reaction cross section shift
$\delta\sigma^{\rm R}_{^{22}\text{C},^{12}\text{C}}$
as a function of the neutron radius of $^{22}$C. }
\label{c22}
\end{figure}

Figure \ref{c22} shows the reaction cross section shift $\delta\sigma^{\rm R}_{22,12}$
as a function of the neutron radius of $^{22}$C. The harmonic oscillator wave function is assumed for $^{22}$C, with
the neutron configuration for the ground states being $(0s_{1/2})^{2}(0p_{3/2})^{4}(0p_{1/2})^{2}(0d_{5/2})^{6}(1s_{1/2})^{2}$.
From the figure, the neutron radius of $^{22}$C is found to be  5.2$\pm$ 0.95 fm.
Using this value and assuming that the proton radius of $^{22}$C
is $2.33$ fm (same as $^{12}$C),
the matter radius of $^{22}$C is found to be $4.59$ fm which is consistent with that of Ref.~\cite{tanaka10}.
Consequently, the neutron skin thickness of $^{22}$C is evaluated to be $2.87\pm 0.95$ fm.
The authors of ~\cite{nagahisa} tried to simultaneously
reproduce the two measured cross sections
of $^{22}$C given in ~\cite{tanaka10} and ~\cite{togano16}, they concluded
that the simultaneous reproduction of the two measured cross sections
of $^{22}$C is not feasible.

As a second application,  the neutron radius
of $^{14}$Be will be determined.
The $^{14}$Be nucleus is also known as a two-neutron halo type
with the two-neutron separation energy  1.27$\pm$0.13 MeV~\cite{wang}.
The interaction cross sections of $^{14}$Be on Carbon and proton targets have been
measured and the matter radius was deduced as
3.22$\pm$0.07 fm~\cite{moriguchi14}, 3.25$\pm$0.11 fm~\cite{Ilieva12},
and 3.1$\pm$0.15 fm~\cite{suzuki99}.
Also, the charge-changing cross section of $^{14}$Be has been measured at 900A MeV on a Carbon
target and the proton radius has been deduced to be 2.41$\pm$0.06~\cite{Terashima14}.

\begin{figure}[t]
\hspace{-.5cm}\epsfig{file=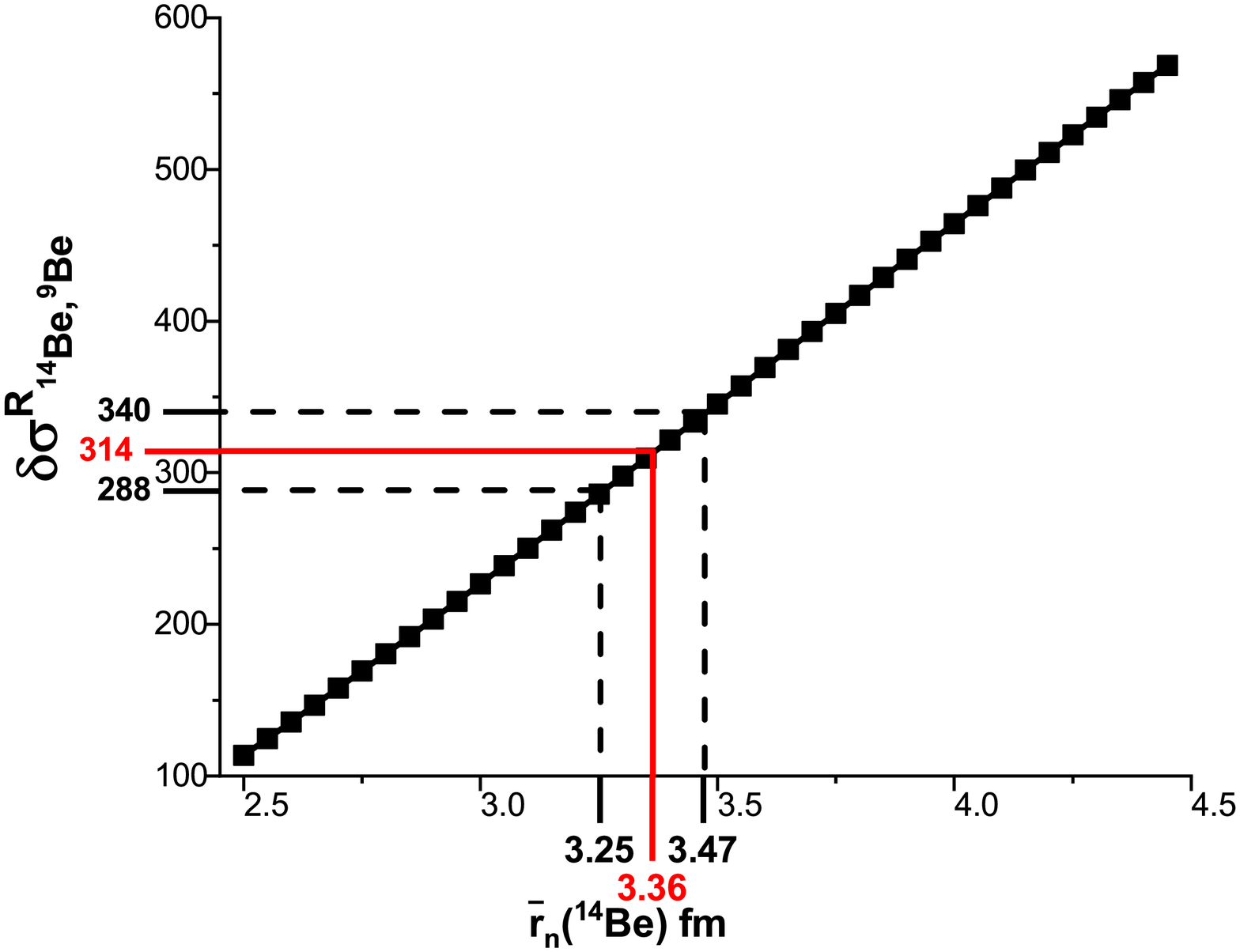,scale=0.3}
\caption{The reaction cross section shift
$\delta\sigma^{\rm R}_{^{14}\text{Be},^{9}\text{Be}}$
as a function of the neutron radius of $^{14}$Be.}
\label{Be14}
\end{figure}

The reaction cross section of $^{14}$Be on a proton target at 41$A$ MeV has been measured
to be $712\pm 14$ mb \cite{moriguchi14}.
On the other hand, the value of the reaction cross section of $^{9}$Be on
a proton target at 39.7$A$ MeV
has been measured to be $398\pm12$ mb \cite{Mcgill74}.
Thus, the reaction cross section shift for $^{14}$Be and $^{9}$Be incident
on a proton target at $\simeq$ 40$A$ MeV is $\delta\sigma^{\rm R}_{14,9}$[Exp]=314$\pm$26 mb.
Assuming a HO wave function for the ground state of $^{14}$Be with neutron configuration
 $(0s_{1/2})^{2}(0p_{3/2})^{4}(0p_{1/2})^{2}(1s_{1/2})^{2}$,
the result of the calculations is shown in Fig.~\ref{Be14}.
From the figure,  the neutron radius of $^{14}$Be is found to be 3.36$\pm$0.11 fm, 
consistent with the results of Ref.~\cite{enyo15,lataz04}.
Assuming that the proton radius of $^{14}$Be is equal to that of $^{9}$Be, i.e. 2.36 fm \cite{ahmed17},
one gets  for the matter radius of $^{14}$Be the value 3.11 fm.
On the other hand, using the proton radius of $^{14}$Be  obtained
from the charge-changing cross section, 2.41$\pm$0.06~\cite{Terashima14},
one gets  the matter radius of $^{14}$Be to be 3.12 fm.
Both agree well with the radii measured in
Refs.~\cite{moriguchi14,Ilieva12,suzuki99}.
The neutron skin thickness of $^{14}$Be  is found  to be
(3.36$\pm$0.11- 2.41$\pm$0.06) $0.95\pm 0.17$ fm.

In summary,  a new method has been introduced to
determine the neutron radii
of neutron-rich nuclei. This method requires
measuring the difference between the reaction cross sections of the neutron-rich
nucleus and that of its stable isotope
on proton targets at the same energy.
By applying this method, the neutron radii
of $^{22}$C and $^{14}$Be have been determined. The obtained radii
are consistent with the previously published values.


\end{document}